# Nonlinear phononics: A new ultrafast route to lattice control

M. Först[1*], C. Manzoni[1#], S. Kaiser[1], Y. Tomioka[2], Y. Tokura[3], R. Merlin[4], and A. Cavalleri[1*]

1) Max-Planck Research Group for Structural Dynamics, University of Hamburg, Center for Free Electron Laser Science, 22607 Hamburg, Germany

2) Correlated Electron Engineering Group, AIST, Tsukuba, Ibaraki, 305-8562 Japan

3) Department of Applied Physics, University of Tokyo, Tokyo, 113-8656 Japan

4) Department of Physics, University of Michigan, Ann Arbor, Michigan 48109-1040, USA

\* Corresponding authors: *michael.foerst@mpsd.cfel.de , andrea.cavalleri@mpsd.cfel.de*

\# Present address: CNR-IFN Dipartimento di Fisica, Politecnico di Milano, 20133 Milan, Italy

**To date, two types of coupling between electromagnetic radiation and a crystal lattice have been identified experimentally. One is direct, for infrared (IR)-active vibrations that carry an electric dipole. The second is indirect, it occurs through excitation of the electronic system and via electron-phonon coupling, as in stimulated Raman scattering [1,2,3]. Nearly 40 years ago, proposals were made of a third path, referred to as ionic Raman scattering (IRS) [4,5]. It was posited that excitation of an IR-active phonon could serve as the intermediate state for Raman scattering, relying on lattice anharmonicities as opposed to electron-phonon interaction [6]. In this paper, we report an experimental demonstration of IRS using femtosecond excitation and coherent detection of the lattice response, by which means we show that this mechanism is relevant to ultrafast optical control in solids. The key insight is that a rectified phonon field can exert a directional force onto the crystal, inducing an abrupt displacement of the atoms from the equilibrium positions. IRS opens up a new direction for the optical control of solids in their electronic ground state [7,8,9], different from carrier excitation [10,11,12,13,14].**

Crystal lattices respond to mid-infrared radiation with oscillatory ionic motions along the eigenvector of the resonantly excited vibration. Let $Q_{IR}$ be the normal coordinate, $P_{IR}$ the conjugate momentum and $\Omega_{IR}$ the frequency of the relevant IR-active mode, which we assume to be non-degenerate, and $H_{IR} = N\left(P_{IR}^2 + \Omega_{IR}^2 Q_{IR}^2\right)/2$ its associated lattice energy ($N$ is the number of cells). For pulses that are short compared to the many-picoseconds decay time of zone-center optical phonons [15] one can ignore dissipation, and the equation of motion is

$$\ddot{Q}_{IR} + \Omega_{IR}^2 Q_{IR} = \frac{e^* E_0}{\sqrt{M_{IR}}} \sin(\Omega_{IR} t) F(t) \tag{1}$$

where $e^*$ is the effective charge, $M_{IR}$ is the reduced mass of the mode, $E_0$ is the amplitude of the electric field of the pulse and $F$ is the pulse envelope. At times much longer than the pulse width

$$Q_{IR}(t) = \left[\int_{-\infty}^{+\infty} F(\tau)d\tau\right] \frac{e^* E_0}{\Omega_{IR}\sqrt{M_{IR}}} \cos(\Omega_{IR} t) \ . \qquad (2)$$

For IRS, the coupling of the IR-active mode to Raman-active modes is described by the Hamiltonian $H_A = -NAQ_{IR}^2 Q_{RS}$ where $A$ is an anharmonic constant and $Q_{RS}$ is the coordinate of a Raman-active mode, of frequency $\Omega_{RS}$, which is also taken to be non-degenerate. Thus, the equation of motion for the Raman mode is

$$\ddot{Q}_{RS} + \Omega_{RS}^2 Q_{RS} = AQ_{IR}^2 \ . \qquad (3)$$

Ignoring phonon field depletion, it follows from Eq. (2) that excitation of the IR mode leads to a constant force on the Raman mode which, for $\Omega_{IR} \gg \Omega_{RS}$, undergoes oscillations of the form

$$Q_{RS}(t) = \frac{A}{2\Omega_{RS}^2}\left[\int_{-\infty}^{+\infty} F(\tau)d\tau\right]^2 \frac{(e^* E_0)^2}{M_{IR}\Omega_{IR}^2}(1 - \cos\Omega_{RS} t) \qquad (4)$$

around a new equilibrium position. Hence, the coherent nonlinear response of the lattice results in rectification of the IR vibrational field and in the concomitant excitation of a lower-frequency Raman-active mode.

We stress that equation (3) describes a fundamentally different process from conventional stimulated Raman scattering [16,17,18], for which the driving term $\hat{\Xi}$ in the equation of motion $\ddot{Q}_{RS} + \Omega_{RS}^2 Q_{RS} = \langle\hat{\Xi}\rangle$ depends only on electron variables [19].

To date, optical nonlinearities from IR-active phonons have only been evidenced in resonantly enhanced second harmonic generation experiments [20,21], while IRS has never been demonstrated experimentally.

Ultrafast optical experiments were performed on single crystal $La_{0.7}Sr_{0.3}MnO_3$, synthesized by the floating zone technique and polished for optical experiments. $La_{0.7}Sr_{0.3}MnO_3$ is a double-exchange ferromagnet with rhombohedrally distorted perovskite structure. Enhanced itinerancy of conducting electrons and relaxation of a Jahn-Teller distortion are observed below the ferromagnetic Curie temperature $T_C$ = 350 K [22,23,24]. Due to the relatively low conductivity, phonon resonances are clearly visible in the infrared spectra at all temperatures [25]. The sample was held at a base temperature of 14 K, in its ferromagnetic phase, and was excited with femtosecond mid-IR pulses tuned between 9 and 19 µm, at fluences up to 2 mJ/cm$^2$. The pulse duration was determined to be 120 fs across the whole spectral range used here. The time-dependent reflectivity was measured with 30-fs pulses at 800-nm wavelength.

Figure 1(a) shows time-resolved reflectivity changes for excitation at 14.3-µm wavelength at 2-mJ/cm$^2$ fluence, resonant with the 75-meV (605 cm$^{-1}$) $E_u$ stretching mode [25,26]. The sample reflectivity decreased during the pump pulse, rapidly relaxing into a long-lived state and exhibiting coherent oscillations at 1.2 THz (40 cm$^{-1}$). This frequency corresponds to one of the $E_g$ Raman modes of $La_{0.7}Sr_{0.3}MnO_3$, which is associated with rotations of the oxygen octahedra [26,27], as sketched in the figure. Consistent with $E_g$ symmetry, we observe a 180-degree shift of the phase of the oscillations for orthogonal probe polarization (Fig. 1(b)).

In contrast, excitation in the near-IR yielded qualitatively different dynamics. A negative reflectivity change of similar size was observed, comparable to what was observed in the ferromagnetic compound $La_{0.6}Sr_{0.4}MnO_3$ [28]. However, only 5.8-THz oscillations were detected, corresponding to the displacive excitation of a 193 -cm$^{-1}$ $A_{1g}$ mode [27,29].

Figure 2(a) shows the time-resolved reflectivity changes for various excitation wavelengths in the mid-IR spectral range. The amplitudes of the initial reflectivity drop, of the long-lived state and of the oscillations all show strong pump wavelength dependence, peaking for excitation near the phonon resonance. In particular, the amplitudes of the 1.2-THz $E_g$ oscillations, plotted in Fig. 2(b), are maximum for excitation near 75-meV, in resonance with the $E_u$ stretching mode. In addition, as shown in Fig. 2(d), we observe a quadratic dependence of the coherent oscillation amplitudes on electric field strength.

These observations are in agreement with the IRS model. According to Eq. (3), the driving force is second order in the mid-IR phonon coordinate, and induces a displacive lattice response analogous to rectification through the second-order susceptibility $\chi^{(2)}$ in nonlinear optics. Thus, one expects the IRS response to peak when the infrared pump field is in resonance with $E_u$ mode, i.e. when $Q_{IR}$ is largest. Secondly, according to Eq. (4) a quadratic dependence of the coherent $E_g$ oscillation amplitude on the mid-IR electric field is expected.

Symmetry considerations are also supporting our interpretation. La$_{0.7}$Sr$_{0.3}$MnO$_3$ crystallizes in the distorted perovskite structure of point group $D_{3d}^6$ (space group $R\bar{3}c$). As mentioned above, the representation of the resonantly driven stretching mode is $E_u$, while the Raman mode is of $E_g$ symmetry. Since $E_g \subset E_u \otimes E_u$, one can write an interaction term of the invariant form

$$H_A = -NA\left[ Q_1^{E_g} Q_x^{E_u} Q_y^{E_u} + Q_2^{E_g} (Q_x^{E_u} Q_x^{E_u} - Q_y^{E_u} Q_y^{E_u}) \right] \tag{5}$$

as required for ionic Raman scattering.

A second experimental observation substantiates our assignment. By using mid-IR pulses in which the carrier-envelope phase offset is stable, we could excite the lattice with a reproducible electric-field phase. To this end, we developed an actively stabilized mid-IR light source based on difference-frequency mixing between two different optical parametric amplifiers [30].

Figure 3 shows the time-resolved reflectivity rise alongside the carrier-envelope phase-stable pump field, as measured in situ by electro-optic sampling in a 50 μm thick GaSe crystal. The time dependent reflectivity shows no signature of the absolute electric-field phase, an effect that is well understood for a driving force resulting from rectification of the lattice polarization.

In summary, we have shown that ionic Raman scattering can be used to control crystal structures in a new way, opening the way to selective lattice modifications impossible with electronic excitations. For example, the nonlinear lattice rectification mechanism could be extended to difference-frequency generation between pairs of non-degenerate excitations, opening up new avenues for the control of condensed matter with light beyond linear lattice excitation.

19. For excitation of insulators below the gap

$$\langle \Xi \rangle \approx \sum_{kl} \frac{\partial \chi_{kl}}{\partial Q_{RS}} V_C E_k(t) E_l(t),$$

where $\partial \chi_{kl} / \partial Q_{RS}$ is the Raman tensor and $V_C$ is the cell volume. For a fully symmetric mode, the corresponding equation of motion is

$$\ddot{Q}_{RS} + \Omega_{RS}^2 Q_{RS} = (\partial \chi / \partial Q_{RS}) V_C \left[ E_0 \sin(\omega_L t) F(t) \right]^2$$

where $\omega_L$ (>> $\Omega_{RS}$) is the laser carrier-frequency. For $t \to \infty$

$$Q_{RS}(t) = \left[ \int_{-\infty}^{+\infty} F^2(\tau) d\tau \right] \frac{V_C E_0^2}{2\Omega_{RS}} \frac{\partial \chi}{\partial Q_{RS}} \sin(\Omega_{RS} t) \qquad . \qquad (5)$$

Thus, the ratio between ionic, $\sigma_I$, and electronic-mediated Raman Scattering cross section, $\sigma_E$ can be estimated as

$$\sigma_I / \sigma_E \sim \frac{e^{*2}}{M_{IR} \Omega_{RS}^2 V_C} \frac{\gamma_G}{\partial \chi / \partial V_C}$$

where $\gamma_G = \partial \ln \Omega_{RS}/\partial V_C$ is the Grüneisen parameter. Since $\gamma_G \sim \partial\chi/\partial V_C \sim V_C^{-1}$, the two cross sections are of similar magnitude.

# ACKNOWLEDGMENT

Work supported in part by the US Air Force Office of Scientific Research under contract FA 9550-08-01-0340 through the Multidisciplinary University Research Initiative Program.

FIGURE CAPTIONS

**Fig. 1 Mid-IR vs. near-IR excitation.** (a) Time-resolved reflectivity changes of $La_{0.7}Sr_{0.3}MnO_3$ detected at the central wavelength of 800 nm for mid-IR excitation at 14.3 μm and near-IR excitation at 1.5 μm. The inset shows the Fourier transform of the oscillatory signal contributions for different pump wavelengths and the atomic displacements of the corresponding phonon modes. (b) Signal oscillations for mid-IR excitation for both parallel (dots) and perpendicular (circles) orientations between the pump and probe polarization. The sample temperature is 14 K.

**Fig. 2 Resonant enhancement at the vibrational mode.** (a) Differential reflectivity as a function of the central mid-IR pump wavelength in the vicinity of the frequency of the $MnO_6$ stretching vibration, together with signal oscillations extracted from the data. The pump fluence is 1.1 mJ/cm$^2$. (b) Plot of the vibrational amplitude, as derived from an extrapolation of the measured oscillations to zero time delay. The red solid line is a Lorentzian fit to the data. (c) Dependence of the vibrational amplitude on the pump electric field measured on resonance at 14.7 μm.

**Fig. 3 Carrier-envelope phase stable excitation.** Relative change of the sample reflectivity induced by carrier-envelope phase stable mid-IR excitation in resonance with the $E_u$-symmetry stretching vibration (dark blue). The electric field of the pump pulse (red), as measured via electro-optic sampling in a 50 μm thick GaSe crystal, and its calculated envelope are also shown. To increase the temporal resolution, we used as probe an optical parametric amplifier which delivered broadband IR pulses (1.2– 2.2 μm) compressed to 14 fs. The probe light was spectrally filtered around 1.6 μm in front of the detector.

FIGURES

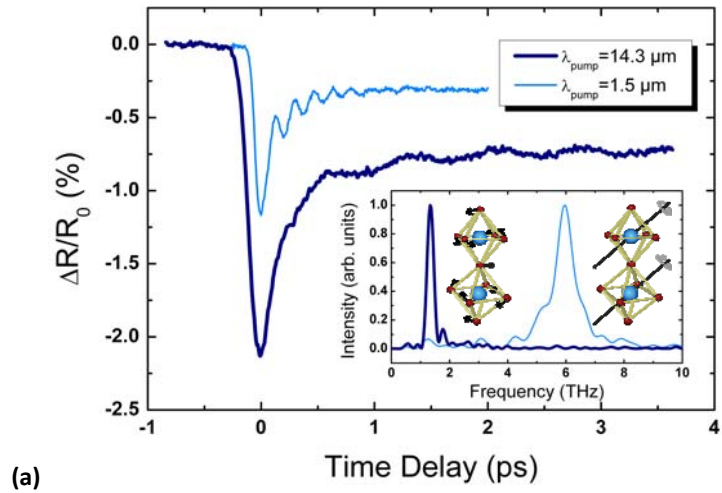

(a)

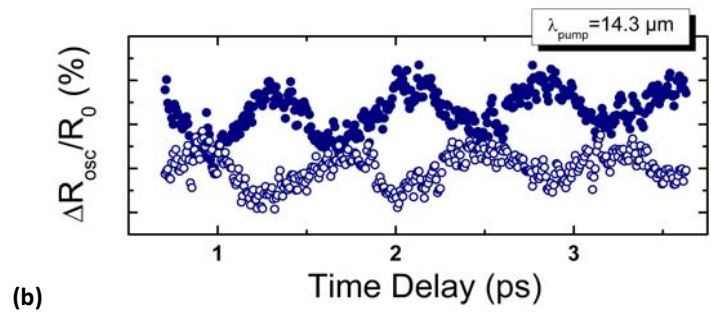

(b)

Först et al., Figure 1

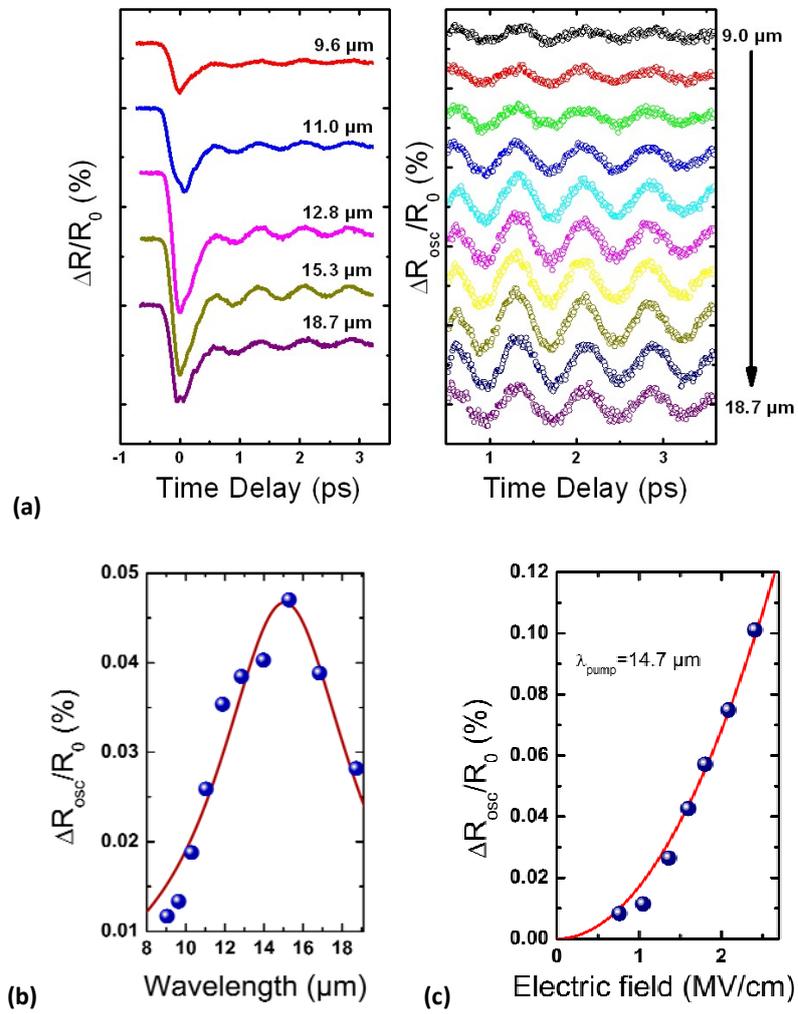

Först et al., Figure 2

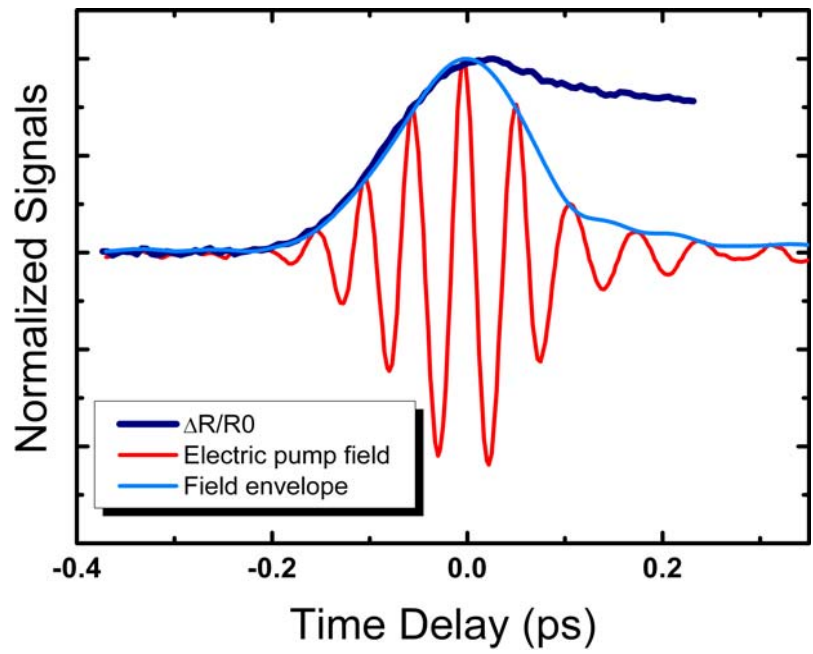

Först et al., Figure 3